\documentclass[journal,draftcls,onecolumn,12pt,twoside]{IEEEtran}
\normalsize
\usepackage{cite}
\usepackage{tabulary}
\usepackage{tabularx}
\usepackage[cmex10]{amsmath}
\usepackage{algorithmic}
\usepackage{graphicx}
\usepackage{textcomp}
\usepackage{amsfonts}
\usepackage{xcolor}
\usepackage{empheq}
\usepackage{float} 
\def\BibTeX{{\rm B\kern-.05em{\sc i\kern-.025em b}\kern-.08em
    T\kern-.1667em\lower.7ex\hbox{E}\kern-.125emX}}

\newcommand{\etal}{\textit{et al}. }

\begin{document}	
\title{Non-Uniform BCSK Modulation in Nutrient-Limited Relay-Assisted Molecular Communication System: Optimization and Performance Evaluation}

\author{Hamid Khoshfekr Rudsari, Mahdi Orooji,~\IEEEmembership{Member,~IEEE},
	Mohammad Reza Javan,~\IEEEmembership{Member,~IEEE},
	Nader Mokari,~\IEEEmembership{Senior Member,~IEEE}, and
	Eduard A. Jorswieck,~\IEEEmembership{Senior Member,~IEEE}
\thanks{Hamid Khoshfekr Rudsari, Mahdi Orooji, and Nader Mokari are with the Department of Electrical and Computer Engineering, Tarbiat Modares University, Tehran 14115111, Iran.}	
\thanks{Mohammad Reza Javan is with Department of Electrical and Robotic Engineering, Shahrood University of Technology, Shahrud 3619995161, Iran.}
\thanks{Eduard A. Jorswieck is with Department of Electrical Engineering and Information Technology, TU Dresden, Germany.}}
\maketitle
\begin{abstract}
In this paper, a novel non-uniform Binary Concentration Shift Keying (BCSK) modulation in the course of molecular communication is introduced. We consider the nutrient limiting as the main reason for avoiding the nanotransmitters to release huge number of molecules at once. The solution of this problem is in the utilization of the BCSK modulation. In this scheme, nanotransmitter releases the information molecules non-uniformly during the time slot. The 3-dimensional diffusion channel with 3-dimensional drift is considered in this paper. To boost the bit error rate (BER) performance, we consider a relay-assisted molecular communication via diffusion. Our computations demonstrate how the pulse shape of BCSK modulation affects the BER, and we also derive the energy consumption of non-uniform BCSK in the closed-form expression.
We study the parameters that can affect the BER performance, in particular the distance between the nanotransmitter and the nanoreceiver, the drift velocity of the medium, and the symbol duration. Furthermore, we propose an optimization problem that is designed to find the optimal symbol duration value that maximizes the number of successful received bits. The proposed algorithm to solve the optimization problem is based on the bisection method. The analytical results show that non-uniform BCSK modulation outperforms uniform BCSK modulation in BER performance, when the aggregate energy is fixed.
\end{abstract}


\section{Introduction} \label{Sec:Introduction}
In nanonetworks, Molecular Communication (MC) is used for communication between nanotransmitters and nanoreceivers. 
The electromagnetic-based communication is not proficient when the communication between microscale and/or nanoscale robots comes into the account \cite{freitas1999nanomedicine}, \cite{sengupta2012fantastic}. Optical communication is also not useful for these microscale and nanoscale applications because the optical signal requires guided medium or line of sight links \cite{farsad2016comprehensive}. This problem is already solved in nature. In the natural world, chemical signals are used for inter- and intra-cellular communication at micro- and nanoscales\cite{bruce2007molecular}. Therefore, chemical signals are a good candidate for communication at both the macro- and microscopic scales. This is the base of MC. It is one of the most competent approach in nanonetworks for communication between nanomachines. In this communication paradigm, nanomachines use molecules to communicate with each other. Moreover, MC signals are biocompatible, and require very low energy, in the order of a femto-Joule (fJ), for transmitting a bit\cite{farsad2016comprehensive}. 

Inspired by the communication methods used by biological systems, a variety of MC frameworks are proposed in the literature \cite{kuran2010energy,enomoto2006molecular,nakano2005molecular,pierobon2010physical}. Among them, microtubule MC is imagined for short-range communication \cite{enomoto2006molecular}, and ion signaling and MC via diffusion are proposed for short to medium range communication \cite{nakano2005molecular,pierobon2010physical}. In this article, we focus on the relay-assisted MC via diffusion (MCvD). It is worth noting that due to practical limitations such as long range distance between transmitter and receiver nodes  and cell signaling in biological communication, relay-aided MC can be adopted. In other words, relays can significantly improve performance and enable the biological communication feasible between different types of cells \cite{einolghozati2014decode,ahmadzadeh2015analysis,atakan2008molecular,nakano2012channel}. 


As a part of the biological systems, the modulation process in MC systems demands the oscillatory behavior in nanotransmitters. This oscillation needs ON/OFF mechanism \cite{chude2019nanosystems}. The nutrient limiting avoid nanotransmitters to synthesize huge number of information molecules (the molecules that are released from the transmitter node to convey the information) at once \cite{polikanov2012hibernation}. This problem, in nature, has been solved by spreading the releasing time of molecules. This is the main idea behind BCSK (Binary Concentration Shift Keying) modulation concept. In standard BCSK, the molecules are released uniformly during the time. Despite uniform BCSK, in this paper, we employ non-uniform pulse shapes in BCSK, and show that the manner of releasing the molecules in time, is going to significantly affect the performance of MC systems.



Different modulation techniques are introduced in MCvD. Mahfuz \etal employ two modulation techniques\cite{mahfuz2010characterization}. In the first technique, they consider transmitting bit ``1" along the concentration of $Q$ (number of) molecules, when for bit ``0'', no molecule is transmitted. In the second modulation technique, they consider information-carrying particles being released as sinusoidal signals with predefined frequency. 
In \cite{kuran2011modulation}, Kuran \etal propose two new modulation techniques, namely Concentration Shift Keying (CSK) and Molecule Shift Keying (MoSK). In CSK, the information is modulated using the concentration of the identical molecules, while different types of molecules are considered in MoSK. This idea is extended to use isomer as messenger molecules by Kim \etal \cite{kim2013novel}.

The channel model considered in this paper is based on 3-dimensional (3-D) diffusion with drift in three dimensions which is studied in \cite{bhatnagar20193}. This model is the best fit with the blood vessel due to the turbulent behavior of the blood flow \cite{chanson2009applied,batchelor1967introduction,felicetti2014molecular}. The relay-assisted MCvD system in 3-D environment is studied in \cite{tiwari2018joint}. In this paper, the authors optimize the location of the relay node, and select the appropriate number of molecules by solving a joint optimization problem. However, the drift of the channel is ignored in the aforementioned paper. The molecular relay channel in the case of 1-D and the capacity of the channel is investigated in \cite{nakano2010design}. In \cite{einolghozati2013relaying}, the authors utilize relaying in MCvD systems in order to enhance the performance of the system in the case of long range distances between the nanomachines. They consider two strategies as sense-and-forward (SF) and decode-and-forward (DF) in the relay node to convey the message to the destination node. In \cite{ahmadzadeh2015analysis}, the authors introduce the scenario of multi-hop relaying in 3-D channel with no drift by evaluating three schemes to design the relay-assisted MCvD system. The schemes are based on different type of molecules and the number of relay nodes. In spite of the fact that the 3-D channel with drift is the most realistic scenario for the channel of MCvD, it has been neglected in the relay assisted channel models. In this paper, the relay assisted MCvD in  the channel with 3-D drift velocity is considered.

In this paper, the bit error rate (BER) of the introduced non-uniform BCSK modulation is derived in closed form expression, and the analytical results reveal that the non uniformity of the molecules concentration shape can potentially improve the BER.
 
In the second part of the paper, the energy model of the MC, when the capacity of the vesicle is not constant, is considered. As the number of the released molecules increases, the BER decreases with the cost of spending more energy. Considering the limited energy per bit, we show the non-uniform BCSK has better BER compared to the uniform BCSK. We also study the trade off between the transmission rate and the probability of successfully receiving a bit. The molecules are transmitting in equal time slots, called the symbol duration. An optimization problem is designed to find the optimum symbol duration by maximizing the number of successful received bits. We propose an algorithm based on the bisection method to find the optimal value of the symbol duration.

The rest of the paper is organized as follows. In Section \ref{Sec:System_model}, we study the system model including the channel model and BER analysis. In this section, we derive the mean and the variance of the number of the received molecules and derive a closed-form expression for bit error probability. In Section \ref{Sec:Energy_model}, we study the energy model to evaluate the BER performance in terms of energy. In Section \ref{Sec:Optimization}, we propose the optimization problem to find the optimal value of symbol duration by utilizing the bisection method. In Section \ref{Sec:Numerical_result}, the numerical result is shown. The paper is concluded in Section \ref{Sec:Conclusion}.

 \textit{Notation}: In this paper, exp($x$) and B($k$,$l$) denote the natural exponential function and the binomial distribution with $k$ total number of experiments and the probability of $l$ that each experiment yielding a successful result, respectively. The notation $\mathcal{N}(\mu , \sigma^2)$ stands for the normal distribution with mean $\mu$ and variance $\sigma^2$. 
 The notations $\text{erf}(x)$ and Pr($X$), refer to error function which is  $ \text{erf} \big(x\big) = \frac{1}{\sqrt{2\pi}} \int_{0}^{x}\text{exp}\big(-y^2\big)\, dy $, and the probability of happening the $X$ event, respectively.

  \begin{figure}[t]
	\centering
	\scalebox{.1}{}
	\includegraphics[width=350pt]{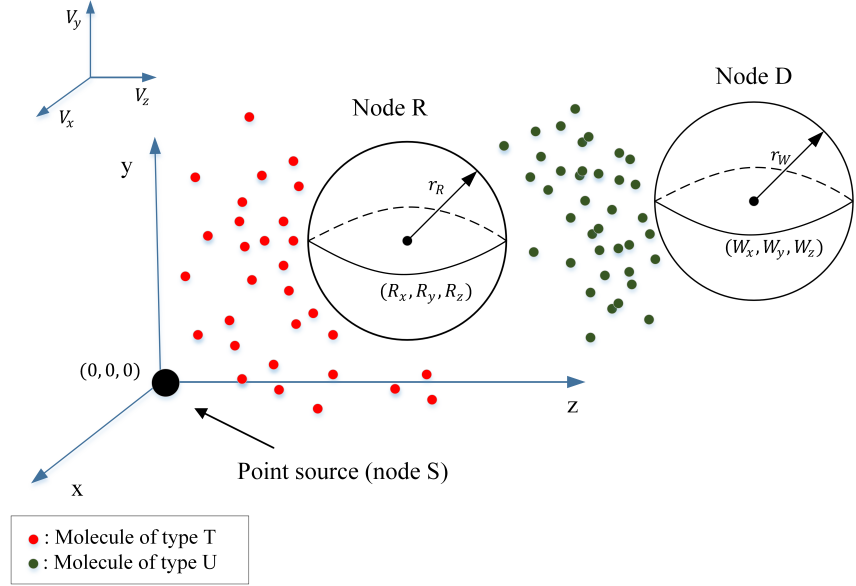}
	\caption{Relay-Assisted MCvD system with drift  }
	\label{Fig:Schematic}
\end{figure}
\section{System Model} \label{Sec:System_model}
In this paper, we adopt Brownian motion \cite{mori1965transport}, to model the movement of the molecules in the medium and consider the drift velocity to attain a comprehensive model for the channel in MCvD system.
	  We employ a generic nanotransmitter as the transmitter node, which can be reused as often as necessary. These types of transmitters are not natural and produced artificially \cite{fakhrullin2012cyborg,chude2019nanosystems}. The receiver in our model is a nanomachine \cite{nakano2014molecular}, which acts as a passive receiver. The passive receiver counts the number of molecules that arrive in its boundary but does not absorb and remove them from the medium \cite{noel2016active}. The nanotransmitter, transmits the information in equal symbol durations. The inter-symbol interference (ISI) is defined as the received molecule during the current symbol duration that is transmitted in the previous symbol durations.
 The noise is created by molecules from the other sources which is accounted in the receiver. The maximum-a-posterior (MAP) rule is employed in the receiver to detect the transmitted information.
 
In the employed model, we have three nodes, namely a nanotransmitter (node S), a nanoreceiver in the destination (node D), and a relay nanomachine (node R). The model is illustrated in Fig.~\ref{Fig:Schematic}.  

This system acts as a relay-assisted MCvD system. Node S is assumed to be a point source. It is also assumed that the nodes do not have any mobility and fixed in their positions. During the transmission node R is located between nodes S and D. 
It has full-duplex transmission and utilizes DF strategy. We assume node S transmits information to node R in the $n^\text{th}$ time slot, node R decodes the received message from node S at the end of the $n^\text{th}$ time slot, and retransmits it to node D in the $(n+1)^\text{th}$ time slot. Finally, node D detects the information at the end of the $(n+1)^\text{th}$ time slot. The medium has drift velocity in three directions, 
and it has uniform temperature and viscosity. 
 
\begin{figure}[t]
	\centering
	\scalebox{.1}{}
	\includegraphics[width=320pt]{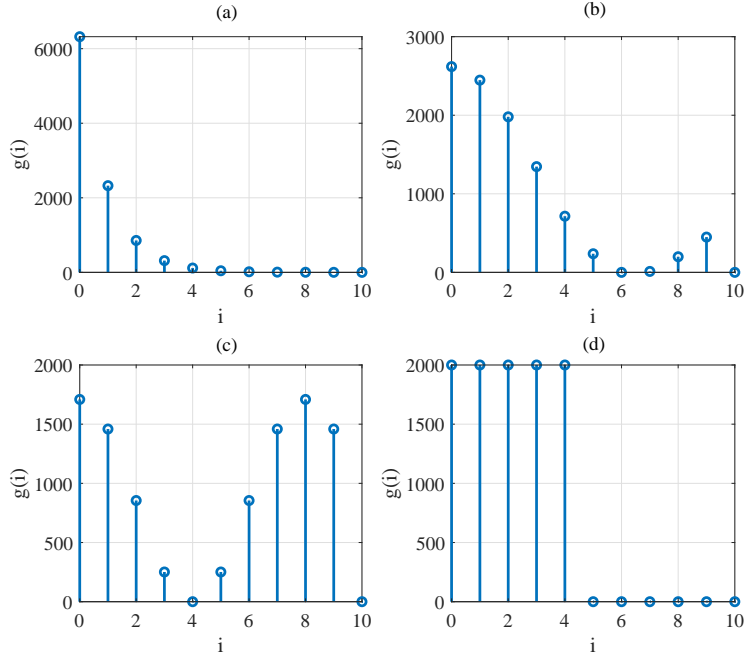}
	\caption{Modulation scheme to send bit ``1'', a). Exponential function for non-uniform BCSK, b). Sinc function for non-uniform BCSK, c). Cosine function for non-uniform BCSK, d). uniform pulse BCSK modulation scheme (the number of molecules to send bit ``1'' in (a), (b), (c) and (d) are equal as $\sum_{i=1}^{10} g(i) = 10000$).}
	\label{Fig:Modulation_scheme_functions}
\end{figure}

 The BCSK modulation technique is based on releasing the same number of molecules in each sub slot of a time slot \cite{yilmaz2017modulation}.
 We propose a new BCSK modulation in which the pulse shape of molecules is non uniform.
We divide each time slot into $I$ identical sub slots, and the time slot duration is denoted by $t_s$. Therefore, each sub slot has $ \big(\frac{1}{I}  \times t_s\big)$ duration. In Fig.~\ref{Fig:Modulation_scheme_functions}, $g(i)$ is the packet of molecules for each sub slot $i$. We should shape $g(i)$ with specific function. In  Fig.~\ref{Fig:Modulation_scheme_functions}, we plot non-uniform BCSK with different pulse shapes. In the introduced scheme, $N$ molecules with the pulse shape of $g(i)$ are released into the medium to represent transmitting bit ``1". In the proposed modulation model, the transmitter does not release any molecule to represent transmitting bit ``0" \cite{mahfuz2010characterization,kuran2011modulation,kim2013novel,garralda2011diffusion}.

  In the relay-assisted MCvD system, we should use two different molecules in order to reduce ISI which are denoted by $T$ and $U$ at the transmitter and relay node, respectively. It is worth noting that the relay node counts the number of molecules of type $T$ and the $U$ molecules does not affect it. In addition, the destination node (node D) counts the number of molecules of type $U$ and is insensitive to molecules of type $T$. 

  \subsection{Channel Model} \label{Subsec:channel_model}
  In this paper, the Brownian motion of the released molecules into the medium is biased in  drift velocity of the medium. 
  The operation of propagating the molecules is administered by both diffusion and drift as a result of the external active mobility. Therefore, we need to consider the Brownian motion with drift in the channel model. We assume no collision occurs between the molecules that are propagating. The released molecules can be diffused with drift in the medium continuously. 
  
  	 We adopt 3-D fluid environment with drift in 3 directions. We aim to consider the probability of arriving the molecules in the volume of the passive spherical receiver which is located at ($W_x$,~$W_y$,~$W_z$) with radius of $r_W$. The relay node is also located between the transmitter and the receiver nodes, at ($R_x$,~$R_y$,~$R_z$) with radius $r_R$. In addition, the point source is located at (0,0,0), and the velocity of the medium is ($V_x$,~$V_y$,~$V_z$). The employed environment is shown in Fig.~\ref{Fig:Schematic}.
  	 
  	 The probability density function (pdf) of the molecules in the 3-D diffusion with drift environment for the point source located at (0,0,0) and a passive spherical receiver with radius  $r_R$ located at ($R_x$,$R_y$,$R_z$) is given by \cite{bhatnagar20193}  	 
  	 \begin{align}
  	 	f(x,y,z,t_s) = \dfrac{1}{\sqrt{(4 \pi D_T t_s)^3}} \exp \bigg( -\dfrac{(x + R_x - V_x t_s)^2 + (y + R_y - V_y t_s)^2 + (z + R_z - V_z t_s)^2 }{4 D_T  t_s}  \bigg), \label{eq:pdf}
  	 \end{align} 	 
   where $ D_T $  is the diffusion coefficient of $T$ molecules in the medium, and $(x, y, z)$ is the location of a point distanced $\sqrt{x^2 + y^2 + z^2}$ from the center of node R. The value of $ D_T $ depends on the temperature\footnote{Body temperature is $\text{310}^\circ$ K \cite{kim2014symbol}.}, viscosity of the fluid\footnote{Viscosity of blood at body temperature is $ 2.46 \times 10^{-3} $ kg/sec.m \cite{kim2014symbol}.}, and the Stokes’ radius of $T$ molecules \cite{tyrrell2013diffusion}.
   
   To achieve the probability of arriving the molecules in the volume of the relay node, the integration of (\ref{eq:pdf}) over the volume of the relay node is required. As studied in \cite{bhatnagar20193}, the CDF of arriving time in (\ref{eq:pdf}), which is the probability of arriving the molecule of type $T$ within time $t_s$ in the volume of the relay node is given by:
   \begin{align}
   	F(t_s)  = \dfrac{r_R^2 }{144 \pi D_T t_s} \exp \bigg(\dfrac{- (R_z - V_z t_s)^2}{4 D_T t_s} \bigg) \cdot \bigg(    4 \sum_{k = 0}^{3} \alpha(k,t_s) +2 \sum_{k=1}^{3} \beta(k,t_s) + 2 \phi(t_s)   \bigg), \label{eq:cdf}
   \end{align}
   where $\alpha(k,t_s)$, $\beta(k,t_s)$, and $\phi(t_s)$ are given as follows \cite{bhatnagar20193}:
   \begin{subequations} \label{eq:sub_equation_CDF}
   	\begin{align}
   	\begin{split} \label{eq:sub_equation_CDF_alpha}
		\alpha(k,t_s) =& \bigg[ \exp\bigg( -\dfrac{ (\frac{2k+1}{8} r_R + R_y - V_y t_s)^2}{4D_T t_s} \bigg)    +     \exp\bigg( -\dfrac{ (-\frac{2k+1}{8} r_R + R_y - V_y t_s)^2}{4D_T t_s} \bigg)\bigg]  \\& \times \bigg[ \text{erf} \bigg( \dfrac{\sqrt{1 - \dfrac{(2k + 1)^2}{64}} r_R - V_x t_s + R_x}{2 \sqrt{D_T t_s}}  \bigg)  -  \text{erf} \bigg( \dfrac{-\sqrt{1 - \dfrac{(2k + 1)^2}{64}} r_R - V_x t_s + R_x}{2 \sqrt{D_T t_s}}  \bigg)  \bigg],
   	\end{split} 
   	\\
   	\begin{split} \label{eq:sub_equation_CDF_beta}
   	\beta(k,t_s) =& \bigg[ \exp\bigg( -\dfrac{ (\frac{k}{4} r_R + R_y - V_y t_s)^2}{4D_T t_s} \bigg)    +     \exp\bigg( -\dfrac{ (-\frac{k}{4} r_R + R_y - V_y t_s)^2}{4D_T t_s} \bigg)\bigg]  \\& \times \bigg[ \text{erf} \bigg( \dfrac{\sqrt{1 - \dfrac{k^2}{16}} r_R - V_x t_s + R_x}{2 \sqrt{D_T t_s}}  \bigg)  -  \text{erf} \bigg( \dfrac{-\sqrt{1 - \dfrac{k^2}{16}} r_R - V_x t_s + R_x}{2 \sqrt{D_T t_s}}  \bigg)  \bigg],
   	\end{split}
   	\\
   	\begin{split} \label{eq:sub_equation_CDF_phi}
   	\phi(t) =& \exp \bigg(- \dfrac{(R_y - V_y t_s)^2}{4 D_T t_s} \bigg) \cdot \bigg[ \text{erf} \big( \dfrac{r_R - V_x t_s + R_x}{2 \sqrt{D_T t_s}} \big) - \text{erf} \big( \dfrac{-r_R - V_x t_s + R_x}{2 \sqrt{D_T t_s}} \big) \bigg].
   	\end{split}  		
   	\end{align}
   \end{subequations}
   The molecular distribution for released molecules from node R toward node D is similar to (\ref{eq:pdf}). The probability of arriving the molecule of type $U$ within time $t_s$ from relay node toward the destination (node D) is given by (\ref{eq:cdf}) and (\ref{eq:sub_equation_CDF}), where $R_x$, $R_y$, and $R_z$ are changed to $W_x - R_x$, $W_y - R_y$, and $W_z - R_z$, respectively. Note that the other parameters in (\ref{eq:cdf}) and (\ref{eq:sub_equation_CDF}) except $t_s$, are changed to the required parameters for the molecules of type $U$.
   
   The CDF in (\ref{eq:cdf}) is a function of the medium velocity, diffusion coefficient of $T$ molecules, the location of node R, and time slot duration. Thus, we can write $ F(t_s) = P_{\text{ar}}(\textbf{V},D_{T},\textbf{R},t_s)$, where $P_{\text{ar}}$ is the probability of arriving $T$ molecules into the volume of the receiver within time $t_s$, $\textbf{V} = (V_x, V_y, V_z)$, and $\textbf{R} = (R_x, R_y, R_z)$.

   In the scenario of multiple relays (more than one relay node) located at $(R_{1k}, R_{2k}, R_{3k})$ for $k = {1, 2, ..., K}$, the value of $(R_x, R_y, R_z)$ in (\ref{eq:cdf}) is replaced with $(R_{1(k-1)} - R_{1(k)}, R_{2(k-1)} - R_{2(k)}, R_{3(k-1)} - R_{3(k)})$. It is worth noting that the investigation of multi-hop MCvD systems is performed in \cite{ahmadzadeh2015analysis}.

    \begin{table}[t]
    	\caption{Specification of Denoted Terms for The Received Molecules}
    	\centering
    	\begin{tabularx}{380pt}{|X|l|}
    		\hline
    		Specification & Denoted Term  \\ \hline
    		\hline
    		The number of molecules transmitted and received during the current time slot & $N_{Cs,r}^T[n]$\\ \hline
    		The number of molecules transmitted in the previous time slot while they are received during the current time slot (ISI) & $N_{Ps,r}^T[n]$ \\ \hline
    		The number of molecules from other sources which act as noise & $N_{Nos,r}[n]$ \\ \hline
    		The error in the counting of the arrived molecules by node R & $N_{NCs,r}[n]$ \\ \hline
    	\end{tabularx}
    \label{table_specifications}
\end{table}
	Let $M_{s,r}^T[n]$ denote the total number of $T$ molecules absorbed by node R and arrived at node R at the end of the $n^\text{th}$ time slot. The total number of molecules, denoted by $M_{s,r}^T[n]$ is given by \cite{singhal2015performance,kuran2011modulation}
	\begin{equation}
	\begin{split}
	M_{s,r}^T[n] = N_{Cs,r}^T[n]+N_{Ps,r}^T[n]+N_{Nos,r}[n] + N_{NCs,r}[n], \label{eq:molecule_distribution}
	\end{split}
	\end{equation}
	where $N_{Cs,r}^T[n]$, $N_{Ps,r}^T[n]$, $N_{Nos,r}[n]$, and $N_{NCs,r}[n]$, are explained at Table~\ref{table_specifications}.

    As regard the independent movement of the messenger molecules with specific probability of arriving the receiver, $N_{Cs,r}^T[n]$ obeys a binomial  distribution as \cite{singhal2015performance,kim2014symbol,yilmaz2015arrival,kuran2011modulation}
    \begin{align}
	N_{Cs,r}^T[n] \sim \sum_{i=0}^{I-1} B(x_{s}[n]  g(i) , P_{1,i} ), \label{binomial}
    \end{align}
     where $x_{s}[n]$ represents the bit transmitted by node S at the $n^\text{th}$ time slot. In (\ref{binomial}) $P_{j,i} = P_{\text{ar}}(\textbf{V},D,\textbf{R},t_{j,i}) $ is the probability of arriving the molecules within time $t_{j,i}$, where $t_{j,i}= \big(jt_s - \frac{it_s}{I}\big)$, is the time of each sub slot $i$ of the $j^\text{th}$ time slot.
    
    The ISI length is finite because we can ignore the ISI effect after a finite number of previously transmitted symbols \cite{kilinc2013receiver}. Therefore, we can write $N_{Ps,r}^T[n]$ as 
\begin{align}
    N_{Ps,r}^T[n] \sim \sum_{j=1}^{J}  \sum_{i=0}^{I-1} B(x_{s}[n-j] \ g(i) , q_{j,i}), \label{ISI_binomial}
    \end{align}
where $J$ indicates the ISI length, $x_{s}[n-j]$ represents the bit transmitted by node S at the $(n-j)^\text{th}$ time slot and $q_{j,i} = P_{j+1,i} - P_{j,i}$.
    
We can approximate the binomial distribution in (\ref{binomial}) and (\ref{ISI_binomial}) by a normal distribution, if $g(i)$ in each sub slot is large enough, and $g(i)P_{\text{ar}}(\textbf{V},D_{T},\textbf{R},t_s)$ is not zero \cite{yilmaz2015arrival}. We know the mean and variance of the sum of independent normal distributions are the sum of their means and variances, respectively. Therefore, $N_{Cs,r}^T[n]$ obeys a normal distribution given by
\begin{align}
       \begin{split}
	 	N_{Cs,r}^T[n]\sim \ \mathcal{N}\bigg( x_{s}[n] \sum_{i=0}^{I-1} g(i) \ P_{1,i}  \ ,x_{s}[n] \sum_{i=0}^{I-1} g(i) \ P_{1,i}(1 - P_{1,i})  \bigg). \label{eq:current_molecule_dist}
       \end{split}
\end{align}
Then, $N_{Ps,r}^T[n]$ becomes
\begin{align}
         \begin{split}
    N_{Ps,r}^T[n] \sim\;  \mathcal{N}\bigg( \sum_{j=1}^{J}  \sum_{i=0}^{I-1}x_{s}[n-j] \ g(i) \ q_{j,i} \ ,\sum_{j=1}^{J}\sum_{i=0}^{I-1}x_{s}[n-j] \ g(i)\ q_{j,i} (1 - q_{j,i})\bigg). \label{eq:previous_molecule_dist}
         \end{split}
\end{align}

We assume the distribution of $N_{Nos,r}[n]$ is normal as \cite{singhal2015performance}
\begin{align}
N_{Nos,r}[n] \sim\; \mathcal{N}(\mu_{nos,r} , \sigma_{nos,r}^2), \label{eq:noise_dist}
\end{align}
where $\mu_{nos,r}$ and $\sigma_{nos,r}^2$ are the mean and variance of noise, respectively. 
The distribution of counting noise $N_{NCs,r}[n]$ is considered as \cite{singhal2015performance}
\begin{equation}
    N_{Ncs,r} [n] \sim\; \mathcal{N}(0 , \sigma_{ncs,r}^2), \label{eq:counter_dist}
\end{equation}
where $\sigma_{ncs,r}^2$ is the variance of counting noise and is dependent on the mean values of the molecules received by node R as (\ref{eq:mu_0}) and (\ref{eq:mu_1}) \cite{singhal2015performance}.
    
Since the distributions of (\ref{eq:current_molecule_dist}), (\ref{eq:previous_molecule_dist}), (\ref{eq:noise_dist}), and (\ref{eq:counter_dist}) are normal, $M_{s,r}^T[n]$ obeys the normal distribution as follows
\begin{align}
&\text{Pr}(M_{s,r}^T[n] \mid x_{s}[n] = 0) \sim \mathcal{N}(\mu_{0s,r},\sigma_{0s,r}^2), \label{eq:molecule_0}
\\
&\text{Pr}(M_{s,r}^T[n] \mid x_{s}[n] = 1) \sim \mathcal{N}(\mu_{1s,r},\sigma_{1s,r}^2), \label{eq:molecule_1}
\end{align}
where $\mu_{0s,r}$, $\sigma_{0s,r}^2$, $\mu_{1s,r}$ and $\sigma_{1s,r}^2$ are derived from (\ref{eq:current_molecule_dist}), (\ref{eq:previous_molecule_dist}), (\ref{eq:noise_dist}), and (\ref{eq:counter_dist}), respectively as follows
\begin{align}
\begin{split}
\mu_{0s,r} = & \ \pi_{1} \sum_{j=1}^{J}\sum_{i=0}^{I-1} g(i) \ q_{j,i} + \ \mu_{nos,r},   \label{eq:mu_0}
\end{split}
\\
\begin{split}
\mu_{1s,r} = & \ \pi_{1} \sum_{j=1}^{J}\sum_{i=0}^{I-1} g(i) \ q_{j,i} + \pi_{1}\sum_{i=0}^{I-1} g(i) \ P_{1,i}  + \mu_{nos,r}, \label{eq:mu_1}
\end{split}
\\
\begin{split}
\sigma_{0s,r}^2 = & \ \pi_{1} \sum_{j=1}^{J} \bigg\{  \sum_{i=0}^{I-1} g(i) q_{j,i}\big(1 - q_{j,i}\big) + \pi_{0} \bigg(\sum_{i=0}^{I-1} g(i) \ q_{j,i}\bigg)^2 \bigg\} + \ \sigma_{nos,r}^2 + \  \mu_{0s,r}, \label{eq:var_0}
\end{split}
\\
\begin{split}
\sigma_{1s,r}^2 = & \ \pi_{1} \sum_{j=1}^{J} \bigg\{  \sum_{i=0}^{I-1}  g(i) q_{j,i}\big(1 - q_{j,i}\big) + \pi_{0} \bigg(\sum_{i=0}^{I-1} g(i) \ q_{j,i}\bigg)^2 \bigg\}  +\sum_{i=0}^{I-1} g(i) \ P_{1,i}\big(1 - P_{1,i}\big) \\&+\sigma_{nos,r}^2+ \mu_{1s,r}, \label{eq:var_1}
\end{split}
\end{align}
where $ \text{Pr}\big(x_{s}[n] = 1\big) = \pi_{1} $  and $ \text{Pr}\big(x_{s}[n] = 0\big) = \pi_{0} $. The details of the calculation for mean and variance of $N_{Ps,r}^T$ are provided in Appendix~\ref{Sec:Appendix}.
  
 Relay-node (node R) decides based on the MAP rule in detection as below \cite{tavakkoli2017performance}
 \begin{equation}
 \begin{split}
 \hat{x}_{r}[n] = \begin{cases} 
 1 & \mbox{if } M_{s,r}^T[n] \mbox{  $ \ge \tau_{R} $} \  \\0 & \mbox{if } M_{s,r}^T[n]  \mbox{  $<\ \tau_{R} $} , \label{eq:molecule_at_receiver}
 \end{cases}
 \end{split}
 \end{equation}
 where $\tau_{R} $ is the detection threshold at node R, and $\hat{x}_{r} [n]$ is the information bit detected by node R at the end of the $n^\text{th}$ time slot. 
 
 Since our system is a relay-assisted MCvD system, node R has to forward the information bit detected at the end of the $n^\text{th}$ time slot. Similar to the mathematical manipulations derived for node S and node R, the distribution of the total number of $U$ molecules arrived at node D, denoted by $M_{r,d}^U[n+1]$, which detected at the end of the $ (n + 1)^\text{th} $ time slot can be derived from 
 \begin{align}
 \begin{split}
 &\text{Pr}(M_{r,d}^U[n+1] \mid x_{r}[n+1] = 0) \sim \mathcal{N}(\mu_{0r,d},\sigma_{0r,d}^2), \label{molecule_relay_0}
 \end{split}
 \\
 \begin{split}
 &\text{Pr}(M_{r,d}^U[n+1] \mid x_{r}[n+1] = 1) \sim \mathcal{N}(\mu_{1r,d},\sigma_{1r,d}^2), \label{molecule_relay_1}
 \end{split}
 \end{align}
 where $x_r[n+1]$ is the transmitted bit at the beginning of the $(n+1)^\text{th}$ time slot from node R. The calculation of mean and variance values of molecules transmitted by node R and received by node D ($\mu_{0r,d}$, $\sigma_{0r,d}^2$, $\mu_{1r,d}$, and $\sigma_{1r,d}^2$), is similar to the calculation of mean and variance values of molecules which are transmitted by node S and received by node R, and are therefore omitted here.

 \subsection{BER Analysis } \label{Subsec:BER_analysis}
We study the BER analysis of the relay-assisted MCvD system, which is introduced in \cite{tavakkoli2017performance}. The probability of error between two nodes for the $n^\text{th}$ transmitted bit is calculated as follows: 
 \begin{equation}
 \begin{split}
 P_{\text{e}}[n]= \ &\text{Pr}(x_{s}[n] = 1)  \text{Pr}\bigg(\hat{x}_{d}(n+1)=0\mid x_{s}[n]=1\bigg) \\&+\text{Pr}(x_{s}[n] = 0)\text{Pr}\bigg(\hat{x}_{d}(n+1)=1\mid x_{s}[n]=0\bigg). \label{eq:error}
 \end{split}
 \end{equation}

 Before the calculation of BER, we should determine the $\text{Pr}(x_{r}[n] = 1)$, which is given by
 \begin{equation}
 \begin{split}
 \text{Pr}(x_{r}[n] = 1) = \ &\pi_{1}\text{Pr}(\hat{x}_{r}[n - 1] = 1 \mid x_{s}[n - 1] = 1) \\ &+ \pi_{0}\text{Pr}(\hat{x}_{r}[n - 1] = 1 \mid x_{s}[n - 1] = 0), \label{eq:prob_in_receiver}
 \end{split}
 \end{equation}
 where $\text{Pr}(x_s[n - 1] = 1)$ and $\text{Pr}(x_s[n - 1] = 0)$ are $\pi_{1}$ and $\pi_{0}$, respectively. We consider the MAP rule detection in (\ref{eq:molecule_at_receiver}) and use (\ref{eq:mu_0}), (\ref{eq:mu_1}), (\ref{eq:var_0}) and (\ref{eq:var_1}) to calculate (\ref{eq:prob_in_receiver}) as follows:
 \begin{equation}
 \begin{split}
 \text{Pr}(\hat{x}_{r}[n - 1] &= 1 \mid x_{s}[n - 1] = 1) =\text{Pr}(M^T_{s,r}[n - 1] \ge \tau_R \mid x_s[n - 1] = 1) \\&=\frac{1}{2}\bigg( 1 - \text{erf}\bigg( \frac{\tau_R - \mu_{1s,r}}{\sqrt{2\sigma_{1s,r}^2}} \bigg) \bigg), \label{eq:conditional_prob_1}
 \end{split}
 \end{equation}
  \begin{equation}
 \begin{split}
 \text{Pr}(\hat{x}_{r}[n - 1] &= 1 \mid x_{s}[n - 1] = 0) =\text{Pr}(M^T_{s,r}[n - 1] \ge \tau_R \mid x_s[n - 1] = 0) \\&=\frac{1}{2}\bigg( 1 - \text{erf}\bigg( \frac{\tau_R - \mu_{0s,r}}{\sqrt{2\sigma_{0s,r}^2}} \bigg) \bigg). \label{eq:conditional_prob_2}
 \end{split}
 \end{equation}
 
Now we can calculate (\ref{eq:error}) from (\ref{eq:conditional_prob_1}) and (\ref{eq:conditional_prob_2}). At the final step we can write BER for the direct transmission from  (\ref{eq:error})-(\ref{eq:conditional_prob_2}) given by
  \begin{equation}
  \begin{split}
  P_{\text{e}_{\text{D}}}[n]=\frac{1}{2}+\frac{1}{4}\bigg[\text{erf}\bigg(\frac{\tau_{D}-\mu_{1s,d}}{\sqrt{2\sigma_{1s,d}^2}}\bigg)-\text{erf}\bigg(\frac{\tau_{D}-\mu_{0s,d}}{\sqrt{2\sigma_{0s,d}^2}}\bigg)\bigg], \label{eq:direct_prob}
  \end{split}
  \end{equation}
  where $\tau_{D}$ is the detection threshold at node D. The BER for the relay-assisted transmission is calculated as follows:
  \begin{equation}
  \begin{split}
  P_{\text{e}_{\text{SR}}}[n]&=\ \frac{1}{2}+\frac{1}{8}\bigg[\text{erf}\bigg(\frac{\tau_{R}-\mu_{1s,r}}{\sqrt{2\sigma_{1s,r}^2}}\bigg)-\text{erf}\bigg(\frac{\tau_{R}-\mu_{0s,r}}{\sqrt{2\sigma_{0s,r}^2}}\bigg)\bigg]\cdot \bigg[\text{erf}\bigg(\frac{\tau_{D}-\mu_{0r,d}}{\sqrt{2\sigma_{0r,d}^2}}\bigg)-\text{erf}\bigg(\frac{\tau_{D}-\mu_{1r,d}}{\sqrt{2\sigma_{1r,d}^2}}\bigg)\bigg], \label{eq:relay_prob}
  \end{split}
  \end{equation}
  where $P_{e_{D}}$ and $P_{e_{SR}}$ denote the bit error probability for the direct transmission and the bit error probability for the relay-assisted transmission, respectively.
  
  
  \section{Energy Model} \label{Sec:Energy_model}
  In a communication via a diffusion system, energy is consumed for the production of the messenger molecules and releasing them toward the destination nanomachine \cite{kuran2010energy}.  In our energy model, the energy cost for sending bit ``1'', is considered as the sum of the total energy of each sub slot $i$ in $g(i)$ pulse shape.
  The steps of the messenger molecule production and releasing known as exocytosis are \cite{bruce2007molecular}:
  \begin{itemize}
  	\item Synthesis of the messenger molecules from their building blocks: The energy cost of a single messenger molecule is denoted by $E_{S}$. Since we should consider the synthesis of $g(i)$ packets in the energy model, the energy cost of it would be $g(i) E_S$.
  	\item Production of a secretory vesicle, which is a packet of the messenger molecules \cite{kuran2010energy}: We use one vesicle for each $g(i)$ packet of the molecules and the energy cost is denoted by $E_{V}(i)$. Since we use one vesicle for each $g(i)$, we would have $C_v = g(i)$ where $C_v$ is the capacity of each vesicle and its unit is the number of molecules.
  	\item Carrying the secretory vesicles to the cell membrane: The energy cost of it denoted by $E_{C}$.
  	\item Releasing the molecules via the fusion of the vesicle and the cell membrane: The energy cost of extraction of a vesicle to the medium denoted by $E_{E}$, which is provided in Table~\ref{table_energy}.
  \end{itemize}
  The capacity of each vesicle $C_{v}$ can be approximated by \cite{kuran2010energy}:
  \begin{equation}
  C_{v} = \bigg(\frac{r_{v}}{r_{mm}\sqrt{3}}\bigg)^3 = g(i), \label{eq:capacity}
  \end{equation}
  where  $r_{v}$ is the radius of a vesicle and $r_{mm}$ is the radius of the messenger molecule.  
  \begin{table}[t]
  	\caption{Energy Model Costs}
  	\centering
  	\begin{tabularx}{360pt}{|X|l|l|}
  		\hline
  		Parameter & Term & Values \\ \hline
  		\hline
  		Energy cost for adding a single amino acid to an amino acid chain &  $E_{AM}$ & $202.88$ \ zJ \cite{kuran2010energy} \\ \hline
  		The cost of synthesizing & $E_{SY}$ & $415$ \ zJ \cite{freitas1999nanomedicine,nelson2008lehninger}\\ \hline
  		The cost of phosphorylation  & $E_{PH}$ & $83$ zJ \cite{freitas1999nanomedicine,nelson2008lehninger} \\ \hline
  		The cost of releasing molecules into the medium  & $E_{E}$ & $830$ \ zJ \cite{eliasson2008novel} \\ \hline
  	\end{tabularx}
  	\label{table_energy}
  \end{table}
  Note that the value of $r_v$ differs for each $g(i)$ due to non-uniform BCSK assumption. Therefore, $r_v$ is not constant in our energy model and we should change $r_v$ to $r_v(i)$ which can be calculated as
\begin{equation}
r_v(i) = \sqrt3 \ r_{mm} \times \sqrt[3]{g(i)}. \label{eq:r_v}
\end{equation}

In this paper, we use proteins as messenger molecules. Proteins are produced by combining amino acids to create a specific amino acid chain at the subunit plant. The energy cost of each step to transmit proteins as messenger molecules that are explained above can be obtained as follows \cite{kuran2010energy}:
\begin{align}
&E_{S} = E_{AM}\big(N_{aa} - 1\big) \ , \label{eq:E_S}\\
&E_{v}(i) = E_{SY}\big(4\pi r_{v}^2(i)\big) , \label{eq:E_v}\\
&E_{C} = E_{PH}\bigg \lceil \frac{r_{unit}/2}{8}\bigg\rceil , \label{eq:E_c}
\end{align}
where $\lceil X \rceil$, $N_{aa}$, $r_{unit}$ are the ceiling function which maps $X$ to the least integer greater than or equal to $X$, the number of amino acids, the radius of the transmitter unit in nanometer scale, respectively. The values of parameters $E_{AM}$, $E_{SY}$, and $E_{PH}$ are provided in Table~\ref{table_energy}. The zJ term in the aforementioned table is the zepto Joule whose value is $10^{-21}$ Joules.

By using all of the energies given above and (\ref{eq:r_v}), the total energy for sending each $g(i)$ is derived as
\begin{align}
\begin{split}
E_{T_i} =  \ E_{AM}\big(N_{aa} - 1\big) g(i) + \bigg(E_{SY}\big(4\pi (\sqrt3 \ r_{mm}  \sqrt[3]{g(i)})^2\big)\ +  \ E_{PH}\bigg \lceil \frac{r_{unit}/2}{8}\bigg\rceil + E_E\bigg). \label{eq:total_energy}
\end{split}
\end{align}

According to (\ref{eq:total_energy}), the total energy for sending bit ``1'' in non-uniform BCSK modulation is
\begin{align}
E_T = \sum_{i=0}^{I-1} E_{T_i}. \label{eq:sum_total_energy}
\end{align}
$E_T$ is directly proportional to pulse shape in non-uniform BCSK modulation as seen in (\ref{eq:total_energy}) and (\ref{eq:sum_total_energy}). 

\section{Symbol Duration Optimization Problem} \label{Sec:Optimization}
 \begin{table}[t]
	\centering
	\caption{Proposed Algorithm Based on The Bisection Method }
	\begin{tabularx}{250pt}{X}
		\hline
		\textbf{Initialization:}\\ 
		Set $ 0 <\epsilon < 1$, $\text{Lower bound}  = 0$, and $\text{Upper bound} = \text{a sufficienctly large number, e.g., $10^{3}$ in our setting}$  \\ \hline \hline
		\textbf{Iterations:}\\ 
		\textbf{Step 1}: $l$ = $(\text{Lower bound} + \text{Upper bound})/2.$\\ 
		\textbf{Step 2}: Solve the concave feasibility problem (\ref{eq:bisection}).\\ 
		\textbf{Step 3}: \textbf{If} (\ref{eq:bisection}) is feasible, \\ \ \ \ \ \ \ \ \ \ \ \ \ $\text{Lower bound} = l$,\\ \ \ \ \ \ \ \ \ \ \textbf{else}\\ \ \ \ \ \ \ \ \ \ \ \ \ $\text{Upper bound} = l$.\\ 
		\textbf{Step 4}: \textbf{If} $\mid \text{Upper bound} - \text{Lower bound} \mid  \ \le \epsilon$, \\ \ \ \ \ \ \ \ \ \ \ \ \ stop, \\ \ \ \ \ \ \ \ \ \ \textbf{else}\\ \ \ \ \ \ \ \ \ \ \ \ \ go back to Step 1.\\ \hline 
	\end{tabularx}
	\label{bisection_method}
\end{table}
In this section we introduce the optimization problem to find the optimal value of the symbol duration.  The transmission data rate, denoted by $R$, is
\begin{align}
R = \frac{1}{t_s}. \label{eq:Rate}
\end{align}
By increasing the symbol duration, the achievable data rate decreases. There is a trade off between $t_s$  and BER, as $t_s$ increases, BER decreases \cite{kadloor2012molecular}. The product of the success probability of a received bit and data rate gives the aforementioned tradeoff. In the binary modulation, for BER more than $0.5$, we can change BER to $1-0.5$ \cite{han2012game}. It is due to the fact that the transmitter node transmits one bit in each time slot, therefore, the probability of successful reception, denoted by $P_{\varUpsilon}[n]$, is \cite{han2012game}
\begin{align}
	P_{\varUpsilon} [n] = (1 - 2 P_{\text{e}_{\text{SR}}}[n])^N, \label{eq:succcess_reception}
\end{align}
where $N$ is the number of bits in each symbol. However, in this paper $N=1$. We formulate the optimization problem to find the optimal value of $t_s^*$ as follows:
\begin{align}
\max_{t_s} \ \ \ \ P_{\varUpsilon} [n](t_s) \cdot R(t_s). \label{eq:OPT}
\end{align}


The objective function in (\ref{eq:OPT}) is quasi-concave, therefore, the optimization problem (\ref{eq:OPT}) is quasi-concave. The discussion of the quasi-concavity of the objective function in (\ref{eq:OPT}) based on the numerical assessments is provided in Appendix~\ref{Sec:Appendix_concave}.

	 We employ the bisection method to solve the optimization problem (\ref{eq:OPT}), in which the optimal symbol duration is determined by solving a concave feasibility problem at each step. The $l$-sublevel set ($l \in \mathbb{R}$) of a function $h : \mathbb{R}^n \to \mathbb{R}$ is defined as \cite{boyd2004convex}
\begin{align}
\zeta_l = \{  x \in \textbf{dom} \ \ h \mid h(x)\ge l \}. \label{eq:sub_level}
\end{align}	
Note that $\mathbb{R}$ is the set of real numbers. Thus, the new optimization problem by considering the $l$-sublevel set of the objective function in (\ref{eq:OPT}) as constraint, is
\begin{align}
\begin{split}
&\text{Find} \ \ \ t_s \\
&\text{s.t}  \ \ \ P_{\varUpsilon} [n](t_s) \cdot R(t_s) - l > 0. \label{eq:bisection}
\end{split}
\end{align}
We provide the proposed algorithm based on the bisection method for solving the optimization problem in Table~\ref{bisection_method}, where $\epsilon$ is some positive small number.

\section{Numerical Result} \label{Sec:Numerical_result}
\begin{table}[]
	\caption{Values and Ranges of MCvD System parameters}
	\centering
	\begin{tabular}{|l|l|l|}
		\hline
		Parameter                          & Variable & Values \\ \hline
		\hline
		Diffusion Coefficient              &      $D_{T}, D_{U}$                     &  $4 \times 10^{-9}$   $ \text{m}^\text{2}$/s \cite{kim2014symbol}\\ \hline
		Drift Velocity                     &       $(V_x, V_y, V_z)$                               &     $[1, 100]$ $\mu$m/s \cite{bhatnagar20193}\\\hline
		Location of node D     &      $(W_X, W_y, W_z)$                           &     [$20$, $200$] $\mu$m \cite{bhatnagar20193}\\ \hline 
		Symbol duration                    &      $t_{s}$                            &    [$1$, $16$] ms \cite{bhatnagar20193}\\ \hline
		Radius of the receiver & $r_R, r_W$ & $50~\mu \text{m}$ \cite{bhatnagar20193} \\ \hline
		Molecular noise mean               &    $\sigma_{nos,r}^2, \sigma_{nor,d}^2$ & $100$ \cite{singhal2015performance}\\ \hline
		Molecular noise variance           &      $\mu_{nos,r}, \mu_{nor,d}$         &$100$ \cite{singhal2015performance}\\ \hline
		Probability of sending bit ``1''       &     $\pi_{1}$                           & $0.5$\\ \hline
		The ISI length                     &    $J$                                  &    $[1, 30]$\\ \hline
		The number of $i$ sub slot                     &    $I$                                  &    $10$\\ \hline
		Number of amino-acids	           &   $N_{aa}$                              &     $51$ \cite{kuran2010energy}\\ \hline
		Stokes' radius of insulin	       &     $r_{s}$                             &   $2.68$ nm  \cite{seki2007effects}\\ \hline
		Radius of insulin molecule 	       &    $r_{mm}$                             &   $2.5$ nm  \cite{freitas1999nanomedicine}\\ \hline
		Radius of the transmitter          & $r_{unit}$                              & $10 \ \mu$m  \cite{bruce2007molecular}\\ \hline
		
	\end{tabular}
	\label{table}
\end{table}
In this section, we show the numerical results to calculate the error probability performance of a relay-assisted MCvD system with drift for BCSK modulation and evaluate the BER in terms of the budget of energy. Also, we present how the MCvD system parameters influence the performance.

The proposed system parameters used in the analysis of BER and energy are provided in Table \ref{table}. 
We also consider protein based messenger molecules in our system. The number of amino acids that are required to form a protein depends on the size of protein. Furthermore, adding a single amino acid to an amino acid chain, regardless of the type of the amino acid, has a constant cost of energy \cite{bermek1997biyofizik}. We assume that the diffusion coefficient of $T$ and $U$ molecules are equal. We also assume $J =$ 10 in the analysis of the system unless otherwise stated. However, the performance of the system is evaluated under the different values of ISI length. 


At first, we assume that node R is located between nodes S and D. Note that node R has the same modulation scheme as node D. It is also assumed that in all figures in this section $W_x = 2 R_x$, $W_y = 2 R_y$, and $W_z = 2 R_z$. Hence, the analysis is based on the location of the relay node.  Without loss of generality, it is assumed that the mean and variance values of molecular noise are considered as 100 \cite{singhal2015performance}. 

\begin{figure}[]
	\centering
	\scalebox{.1}{}
	\includegraphics[width=330pt]{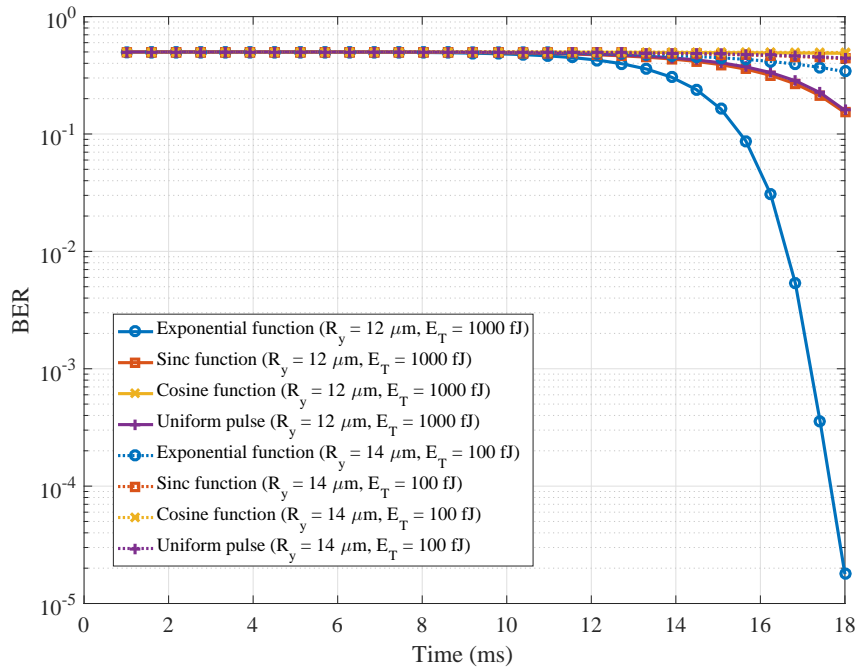}
	\caption{BER of the MCvD system with the relay node for exponential, sinc, cosine and uniform pulse shape in non-uniform BCSK modulation, as a function of symbol duration ($t_s$) for different locations of relay node and different values of energy for transmitting bit ``1'' ($R_x = 100~\mu$m, $R_z = 14~\mu$m, $V_x = 100~\mu$m/s, $V_y = 40~\mu$m/s, and $V_z = 40~\mu$m/s).}
	\label{Fig:Function_of_time}
\end{figure}

The BER performance of MCvD system for exponential, sinc, cosine, and uniform pulse shapes as a function of the symbol duration, $t_s$, are presented in Fig.~\ref{Fig:Function_of_time}. In addition, the analysis is based on two different locations of the relay node, where $\textbf{R} = (\text{100}~\mu \text{m}, \text{12}~\mu \text{m}, \text{14}~\mu \text{m})$ and $\textbf{R} = (\text{100}~\mu \text{m}, \text{14}~\mu \text{m}, \text{14}~\mu \text{m})$. The energy consumption to transmit bit ``1'' is also identical for each different value of the relay nodes location. As shown in this figure, the BER performance improves by increasing the time slot duration. The location of the relay node also impacts the BER performance. The BER of MCvD system for the relay and destination nodes in case they are closer to the transmitter, is smaller. The exponential function outperforms other pulse shapes in the identical time slot, location of the relay, and energy consumption. On the other hand, the BER performance of the sinc function is better compared to uniform pulse and cosine function. The MCvD system reaches the BER around $\text{10}^{-\text{5}}$ in case of utilizing the exponential function, the pulse shape set to $t_s = \text{18}$~ms, the destination node is located at $(\text{200}~\mu \text{m}, \text{24}~\mu \text{m}, \text{28}~\mu \text{m})$, and the energy consumed to transmit bit ``1'' is 1000~fJ.
\begin{figure}[]
	\centering
	\scalebox{.1}{}
	\includegraphics[width=330pt]{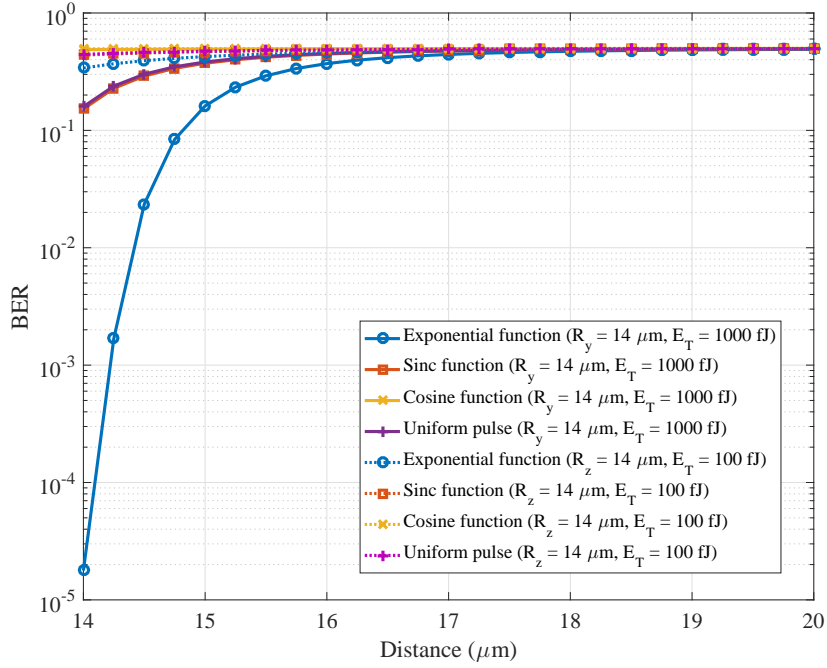}
	\caption{BER of the MCvD system with the relay node for exponential, sinc, cosine and uniform pulse shape in non-uniform BCSK modulation, as a function of location of the relay node for different values of energy for transmitting bit ``1'' ($R_x = 100~\mu$m, $V_x = 100~\mu$m/s, $V_y = 40~\mu$m/s, $V_z = 40~\mu$m/s, and $t_s = 18$~ms).}
	\label{Fig:Function_of_distance}
\end{figure}

In Fig.~\ref{Fig:Function_of_distance}, we present the BER performance of non-uniform BCSK modulation as a function of the location of the node R. It is assumed that $R_x = \text{100}~\mu \text{m}$. We can observe that by increasing $R_z$, for fix value of $R_y = \text{14}~\mu \text{m}$, the BER increases. It is due to the fact that by increasing the distance between the transmitter and the receiver nodes, the probability of hitting the molecules by the receiver decreases. In addition, the BER performance of the relay-assisted MCvD system for fixed value of $R_z = \text{14}~\mu \text{m}$ and as a function of $R_y$ is shown in this figure. By increasing $R_y$, BER is also increased. Also, as shown in Fig.~\ref{Fig:Function_of_distance}, the exponential pulse shape for non-uniform BCSK modulation has the best BER performance among other examples. It is worth noting that the analysis is based on of two different values of energy consumption. It can be viewed from the Fig.~\ref{Fig:Function_of_distance} that the more consumed energy, the less the value of BER is. The BER of the proposed MCvD system changes from $\text{10}^{-\text{5}}$ to around $\text{10}^{-\text{3}}$ by increasing $R_z$ from $\text{14}~\mu \text{m}$ to $\text{14.25}~\mu \text{m}$ in case of considering $R_y = \text{14}~\mu \text{m}$ and exponential function as the pulse shape of the system. Hence, the performance of the system is extremely sensitive to the locations of the relay and destination node.

\begin{figure}[t]
	\centering
	\scalebox{.1}{}
	\includegraphics[width=330pt]{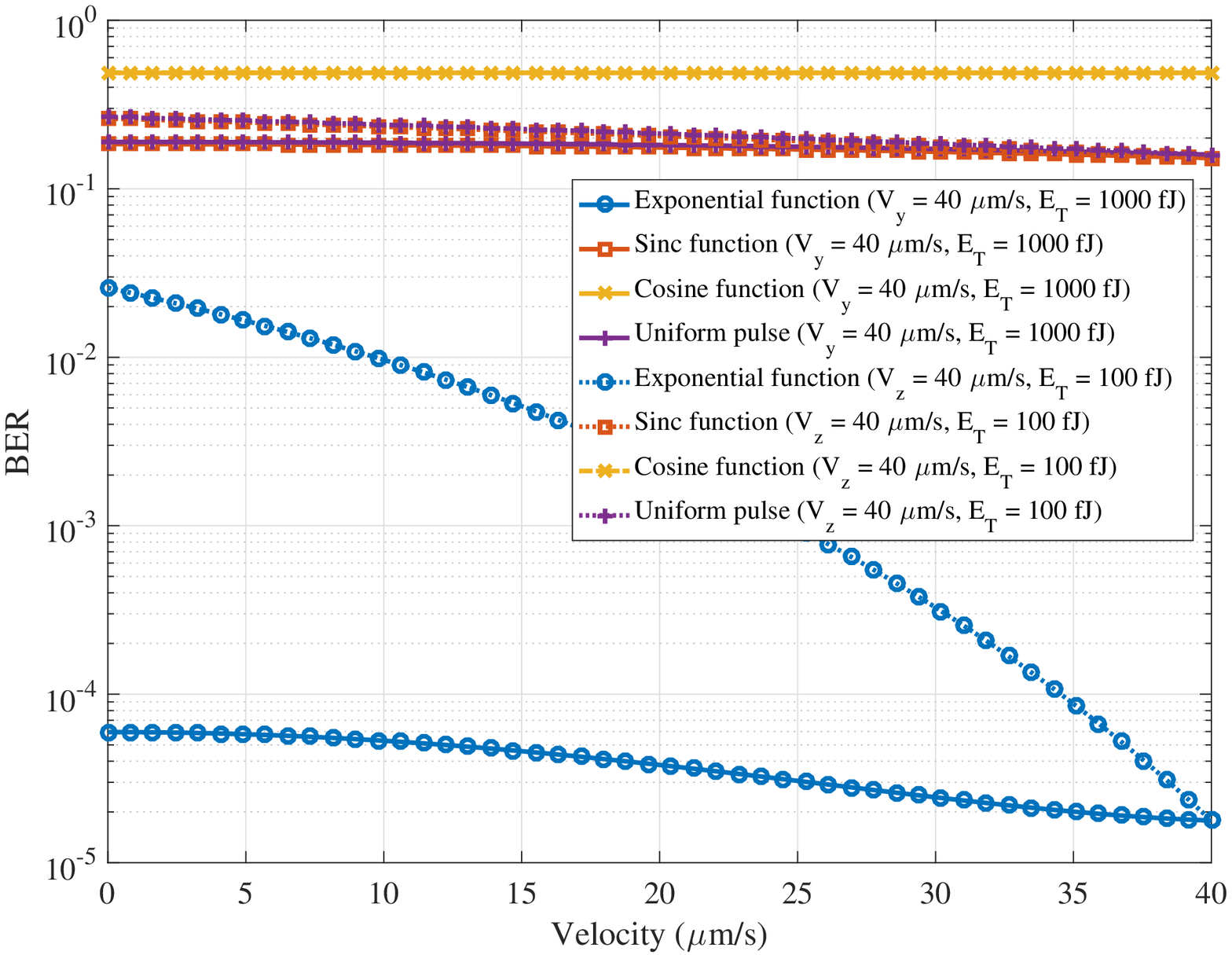}
	\caption{BER of the MCvD system with the relay node for exponential, sinc, cosine and uniform pulse shape in non-uniform BCSK modulation, as a function of drift velocity for different values of energy for transmitting bit ``1'' ($R_x = 100~\mu$m, $R_y = 12~\mu$m, $R_z = 14$ $\mu$m, $V_x = 100~\mu$m/s, and $t_s = 18$~ms).}
	\label{Fig:Function_of_velocity}
\end{figure}

\begin{figure}[]
	\centering
	\scalebox{.1}{}
	\includegraphics[width=330pt]{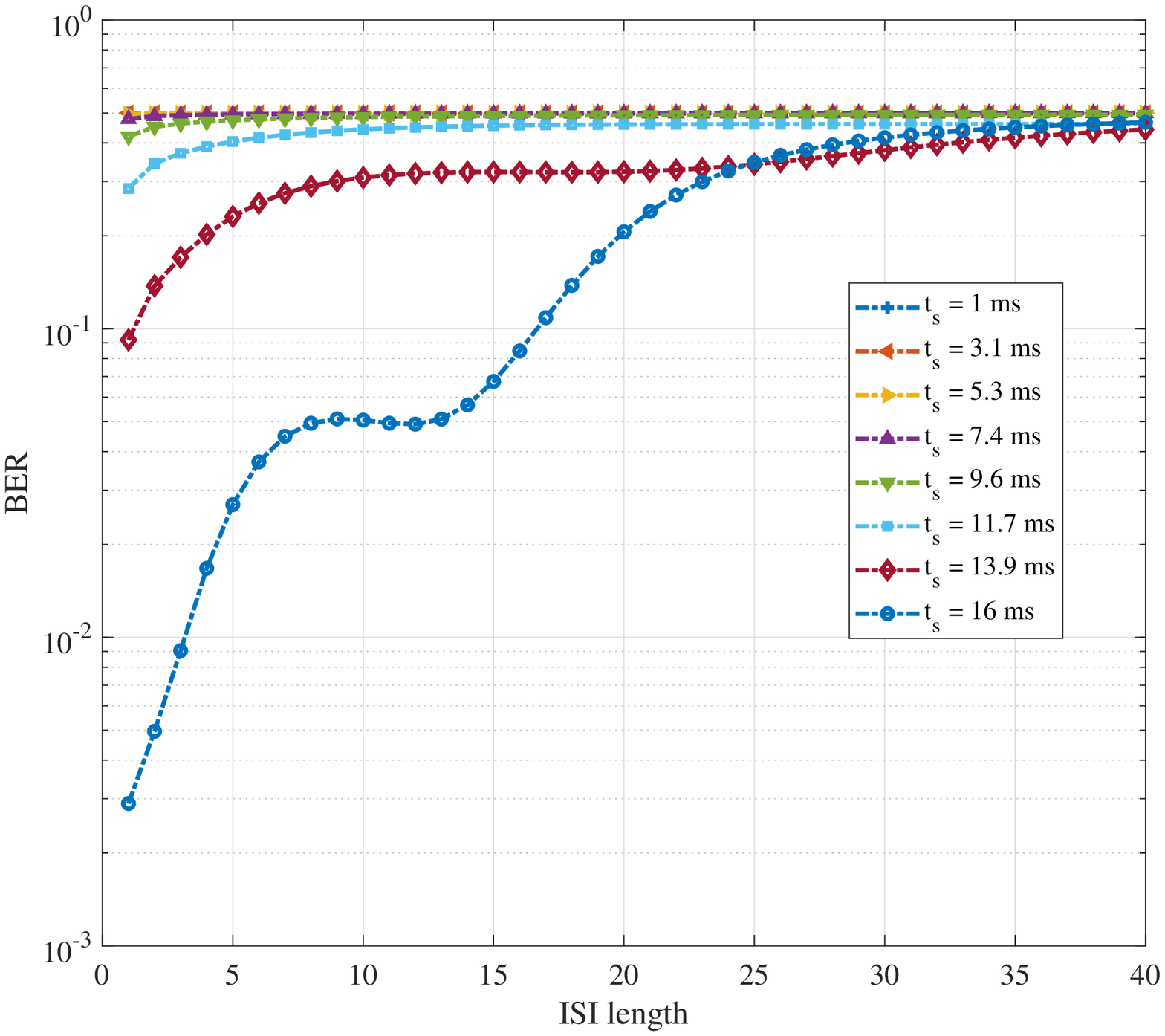}
	\caption{BER of the MCvD system with the relay node for exponentially function pulse shape in non-uniform BCSK modulation, as a function of ISI length for different values of symbol duration ($R_x = 100~\mu$m, $R_y = 12~\mu$m, $R_z = 14$ $\mu$m, $V_x = 100~\mu$m/s, $V_y = 40~\mu$m/s, and $V_z = 40~\mu$m/s).}
	\label{Fig:Function_of_ISI2}
\end{figure} 
In Fig.~\ref{Fig:Function_of_velocity}, the BER performance of different pulse shapes for non-uniform BCSK modulation, fixed value of $V_x = \text{100}~\mu \text{m}$, and different values of drift velocity of $V_y$ and $V_z$ is depicted. It shows that by increasing $V_z$ from 0 to $\text{40}~\mu \text{m/s}$ and for $V_y = \text{40}~\mu \text{m/s}$ and fixed value of consumed energy as $E_T = \text{1000~fJ}$, BER of the system decreases from $\text{6} \times \text{10}^\text{-5}$ to $\text{1.7} \times \text{10}^\text{-5}$ in case of exponential function as the pulse shape. In the scenario of utilizing sinc function as the pulse shape, by considering the mentioned setting of the system parameters, BER decreases from 0.1841 to 0.1522, which is smaller compared to the exponential function for pulse shaping. On the other hand, by assuming the fixed value of $V_z = \text{40}~\mu \text{m/s}$ and different values of $V_y$ from 0 to  $\text{40}~\mu \text{m/s}$, the BER of uniform pulse shape decreases from 0.267 to 0.1592. On the other hand, in case of utilizing the exponential function, BER decreases from 0.0258 to $\text{1.7} \times \text{10}^\text{-5}$, which has the best performance among the other pulse shapes with the aggregated consumed energy is 100 fJ. By taking the aforementioned explanation into consideration, increasing the drift velocity improves the BER performance of the system in fixed energy consumed to transmit bit a ``1''. We can see that increasing the drift velocity can improve the BER performance, because the drift velocity can make the diffusion and propagating molecules faster. Thus, increasing the drift velocity can improve the arriving probability of molecules which results in the BER performance improvement. Increasing the drift velocity also impacts the number of molecules received from other sources, which are denoted by $N_{Nos,r}[n]$ and treated as noise. Furthermore, BER is improved by increasing the drift velocity as shown in Fig.~\ref{Fig:Function_of_velocity}, which implies that the impact of received molecules during the current time slot overcomes the noise and ISI effects.

The length of ISI impacts the BER performance \cite{kilinc2013receiver}. In Fig.~\ref{Fig:Function_of_ISI2}, the BER performance for the exponential function as the pulse shape at the transmitter, as a function of ISI length from 0 to 40, for different values of time slot is shown. It could be seen that by increasing the ISI length, the BER increases. It is due to the fact that the molecules transmitted in the previous time slots are received in the current time slot. By increasing the value of $t_s$, the BER decreases, because the molecules have enough time to arrive at the relay node, and after that to destination node. It also could be find that by increasing the time slot duration, the previously transmitted molecules have more time to arrive at the receiver in their time slot, therefore, the BER decreases. By considering the ISI length from 6 to 14 for $t_s = \text{16~ms}$, the BER increases from 0.037 to 0.0566. However, for ISI length greater than 17, the BER goes over 0.1, for which the impact of ISI could be ignored. 

\begin{figure}[]
	\centering
	\scalebox{.1}{}
	\includegraphics[width=330pt]{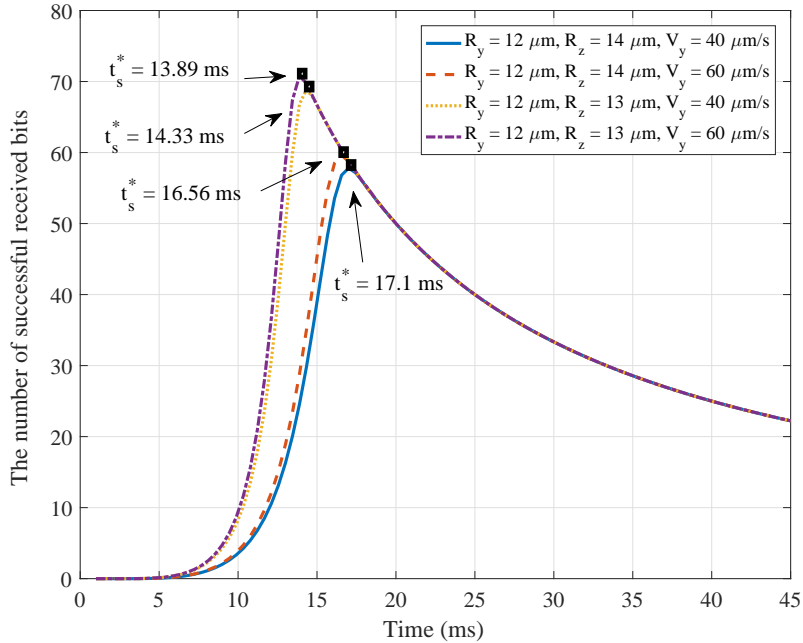}
	\caption{The optimization evaluation based on the proposed algorithm to find the optimal symbol duration $t_s$ ($R_x = 100~\mu$m, $V_x = 100~\mu$m/s, and $V_z = 40~\mu$m/s).}
	\label{Fig:Optimization}
\end{figure}

The performance of the proposed algorithm to obtain the optimal value of the symbol duration is demonstrated in Fig~\ref{Fig:Optimization}. We consider four cases for the mixture of the location of the relay node and $V_y$ when the pulse shape is exponential function. As we discussed, the optimal symbol duration obtains the maximum number of successful received bits. By considering $V_y = \text{60}~\mu \text{m/s}$, the location of the relay node as $(\text{100}~\mu \text{m}, \text{12}~\mu \text{m}, \text{13}~\mu \text{m})$, and optimal symbol duration as 13.89~ms, the number of successful received bits reaches to $\text{70.8}$. We observe that by increasing the velocity of the medium in fixed location of the relay node, the optimal symbol duration is decreased. This is due to the Brownian motions behavior. Increasing drift velocity causes faster propagation of the molecules into the medium. In addition, the number of molecules that are received by the receiver node are increased. Therefore, we can find the smaller value of the symbol duration to reach the minimum BER. It also could be realized that for nearest location of the relay node and the destination node, the number of successful received bits increases. On the other hand, the symbol duration decreases. It is due to the fact that, by decreasing the distance between the transmitter and receiver, the molecules need smaller time to arrive at the receiver. In addition, the probability of arriving molecules in short distances, is more than the long ones. Hence, the number of successful received bits increases.
 
Note that we use MAP rule detection in Fig.~\ref{Fig:Function_of_time}, Fig.~\ref{Fig:Function_of_distance}, Fig.~\ref{Fig:Function_of_velocity}, and Fig.~\ref{Fig:Function_of_ISI2}, at receiver node. We can observe from Fig.~\ref{Fig:Function_of_time}, Fig.~\ref{Fig:Function_of_distance}, and Fig.~\ref{Fig:Function_of_velocity}, that the pulse shaping in BCSK modulation can make a significant difference in BER performance considering parameters of MCvD system and budget of energy.

\section{Conclusion} \label{Sec:Conclusion}
The nutrient limiting is the main reason to utilize the BCSK modulation in MC systems. In this paper, we proposed a new non-uniform BCSK modulation and analyzed the BER of a relay-assisted MCvD system in a 3-D environment with drift with respect to the energy in nanotransmitter. 
 We also derived closed-form expression for the means and the variances of the number received molecules to calculate BER in the direct and the relay-assisted transmission.
In addition, we studied the effect of the system parameters such as the distance between nanotransmitter and nanoreceiver, drift velocity of medium and symbol duration on the BER performance. We also employed an iterative algorithm to find the optimal value of symbol duration in terms of maximizing the number of successfully received bits. Finally, we compared the difference between uniform and non-uniform BCSK modulation considering the energy and BER performance. 
As a future work, one can optimize the pulse shape in the introduced modulation model to minimize the BER in a channel with degradation of molecules. The other future works can focus on the outage performance, subject to the evaluation of the Signal-to-Noise-Ratio (SINR) or other goodputs, such as the capacity of the channel.  One can also optimize the other parameters of the system, such as the number of allocated molecules to the transmitter node. Furthermore, interested readers can investigate applications of MC, e.g., the mobile MC and drug delivery system on the basis of the introduced non-uniform BCSK modulation. Position of the relay in relay-assisted MCvD system has a massive impact on the BER performance. To solve this issue, one of our future works is based on the optimization of the location of the relay node based on the minimization of BER or other outputs of the system.

\appendices
\section{The Calculations of Mean and Variance for ISI} \label{Sec:Appendix}
In this appendix, the details for calculation of mean and variance values of $N_{Ps,r}^T$ in (\ref{eq:previous_molecule_dist}), are provided.

We approximate the binomial distribution of each sub slot $i$ in every time slot $j$ to normal distribution in (\ref{eq:previous_molecule_dist}), in terms of $g(i)$ is large enough and $g(i) \times q_{j,i}$ is not zero. The mean of ISI for the $j^\text{th}$ time slice is calculated as follows:
\begin{align}
\begin{split}
E(N_{Ps,r}^T[n,j]) &= \ \pi_{0} E(N_{Ps,r}^T[n,j] \mid x_{s}[n-j] = 0) + \pi_{1} E(N_{Ps,r}^T[n,j] \mid x_{s}[n-j] = 1) \\&= \pi_{1} \sum_{i=0}^{I-1} g(i) q_{j,i}. \label{eq:previous_molecule_expectation}
\end{split}
\end{align}

The variance value of $N_{Ps,r}^T$ for $j^\text{th}$ time slice is calculated as
\begin{align}
\begin{split}
var(N_{Ps,r}^T[n,j]) = E((N_{Ps,r}^T[n,j])^2) - E^2(N_{Ps,r}^T[n,j]), \label{eq:previous_molecule_var}
\end{split}
\end{align}
where $var(X)$ denotes the variance of random variable $X$ and $E((N_{Ps,r}^T[n,j])^2)$ can be calculated as follows:
\begin{align}
\begin{split}
E((N_{Ps,r}^T[n,j])^2) = \ &\pi_{0} E((N_{Ps,r}^T[n,j])^2 \mid x_{s}[n-j] = 0) + \pi_{1} E((N_{Ps,r}^T[n,j])^2 \mid x_{s}[n-j] = 1) \\ = \ & \pi_{1}\bigg( \sum_{i=0}^{I-1} g(i) q_{j,i}(1 - q_{j,i}) + \big(\sum_{i=0}^{I-1} g(i) q_{j,i}\big)^2\bigg). \label{eq:previous_molecule_expectation_square}
\end{split}
\end{align}
Therefore, $var(N_{Ps,r}^T[n,j])$ becomes
\begin{align}
\begin{split}
var(N_{Ps,r}^T[n,j]) =\ & \pi_{1}\bigg( \sum_{i=0}^{I-1} g(i) q_{j,i}(1 - q_{j,i}) 
+ \big(\sum_{i=0}^{I-1} g(i) q_{j,i}\big)^2 \bigg) - \bigg(\pi_{1} \sum_{i=0}^{I-1} g(i) q_{j,i}\bigg)^2 \\ = \ & \pi_{1}\bigg( \sum_{i=0}^{I-1} g(i) q_{j,i}(1 - q_{j,i}) \bigg)+ \pi_{0}\pi_{1}\bigg(\sum_{i=0}^{I-1} g(i) q_{j,i}\bigg)^2.
\end{split} \label{eq:var_of_ISI}
\end{align}

The mean and variance values for ISI are, respectively, derived as
\begin{align}
&E(N_{Ps,r}^T[n]) = \sum_{j=1}^{J} E(N_{Ps,r}^T[n,j]), \label{eq:totsl_mean_ISI} \\
&var(N_{Ps,r}^T[n]) = \sum_{j=1}^{J} var(N_{Ps,r}^T[n,j]). \label{eq:totsl_var_ISI}
\end{align}

\section{The Discussion of the Quasi-concavity of the Proposed Optimization Problem} \label{Sec:Appendix_concave}

\begin{figure}[]
	\centering
	\scalebox{.1}{}
	\includegraphics[width=330pt]{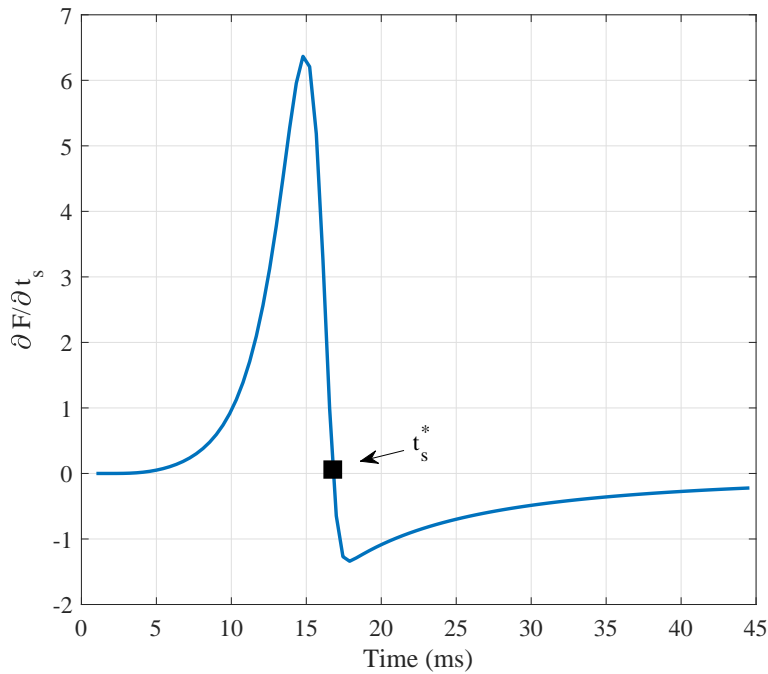}
	\caption{The first derivative of the objective function proposed in (\ref{eq:OPT}) with respect to $t_s$, as a function of time slot duration ($R_x = 100~\mu$m, $R_y = 12~\mu$m, $R_z = 13$ $\mu$m, $V_x = 100~\mu$m/s, $V_y = 40~\mu$m/s, and $V_z = 40~\mu$m/s).}
	\label{Fig:optimization_quasiconcave}
\end{figure}

The derivation of the proposed objective function in (\ref{eq:OPT}) with respect to the optimization variable $t_s$ is not tractable due to the complexity of the BER formula in (\ref{eq:relay_prob}) which is also dependent on (\ref{eq:mu_0})-(\ref{eq:var_1}). By assuming the objective function in (\ref{eq:OPT}) as $F$, we plot the first derivative of the objective function $F$ with respect to $t_s$ in Fig.~\ref{Fig:optimization_quasiconcave} by considering the exponential function for $g(i)$. Note that if we simulate the first derivative of $F$ with respect to $t_s$ for all possible values of $t_s$, the results are always similar to Fig.~\ref{Fig:optimization_quasiconcave}. From out extensive simulations, we deduce that the first derivative of $F$ with respect to $t_s$ is non-negative for the values of $t_s$ up to a point denoted by $t_s^*$ and is non-positive the values above it, i.e., it is non-decreasing for $t_s \le t_s^*$ and is non-increasing for $t_s \ge t_s^*$. Hence, the first derivative of the objective function in (\ref{eq:OPT}) is given by
\begin{subequations} \label{eq:inc_dec}
	\begin{empheq}[left=\empheqlbrace]{align}
	\dfrac{\partial F}{\partial t_s} >0, &\ \ \ \ \   t_s \le t_s^* \label{eq:inc_dec_a}\\
	\dfrac{\partial F}{\partial t_s} <0, &\ \ \ \ \   t_s \ge t_s^*. \label{eq:inc_dec_b}
	\end{empheq}
\end{subequations}

Therefore, (\ref{eq:OPT}) is quasi-concave. 
Finally, due to the quasi-concavity of the objective function, we conclude that our proposed optimization problem is quasi-concave.
\bibliographystyle{IEEEtran}
\bibliography{Hamid_Khoshfekr}

\end{document}